# The role of breccia lenses in regolith generation from the formation of small, simple craters: Application to the Apollo 15 landing site

**M. Hirabayashi[1], B. A. Howl[2], C. I. Fassett[3], J. M. Soderblom[4], D. A. Minton[2], and H. J. Melosh[2].**

[1]Department of Aerospace Engineering, Auburn University, Auburn, AL 36849, U.S.A.

[2]Department of Earth, Atmospheric and Planetary Sciences, Purdue University, West Lafayette, IN 47907, U.S.A.

[3]NASA Marshall Space Flight Center, Huntsville, AL 35805, U.S.A.

[4]Department of Earth, Atmospheric and Planetary Sciences, Massachusetts Institute of Technology, Cambridge, MA 02139, U.S.A.

Corresponding author: Masatoshi Hirabayashi (thirabayashi@auburn.edu)

**Key Points:**

- We develop an analytical expression of regolith generation by small, simple craters.
- This model describes well the statistics of regolith formation, time evolution, and regolith growth.
- Our method also provides consistent results with empirical observations of regolith at the Apollo 15 landing site.

## Abstract

Impact cratering is likely a primary agent of regolith generation on airless bodies. Regolith production via impact cratering has long been a key topic of study since the Apollo era. The evolution of regolith due to impact cratering, however, is not well understood. A better formulation is needed to help quantify the formation mechanism and timescale of regolith evolution. Here, we propose an analytically derived stochastic model that describes the evolution of regolith generated by small, simple craters. We account for ejecta blanketing as well as regolith infilling of the transient crater cavity. Our results show that the regolith infilling plays a key role in producing regolith. Our model demonstrates that, because of the stochastic nature of impact cratering, the regolith thickness varies laterally, which is consistent with earlier work. We apply this analytical model to the regolith evolution at the Apollo 15 site. The regolith thickness is computed considering the observed crater size-frequency distribution of small, simple lunar craters (< 381 m in radius for ejecta blanketing and < 100 m in radius for the regolith infilling). Allowing for some amount of regolith coming from the outside of the area, our result is consistent with an empirical result from the Apollo 15 seismic experiment. Finally, we find that the timescale of regolith growth is longer than that of crater equilibrium, implying that even if crater equilibrium is observed on a cratered surface, it is likely the regolith thickness is still evolving due to additional impact craters.

## Plain Language Summary

Impact cratering likely generates much of the regolith (the surface layer made up of a mixture of rocks, rock fragments, sand, and dust) observed on airless planetary surfaces. However, the way that the regolith layer evolves and thickens over time due to impact cratering events is not well understood. When a small, simple crater forms into hard rock, regolith is produced by fracturing the target rock and is deposited in the crater's ejecta blanket and within its transient crater cavity. Here, we discuss an analytically derived stochastic model that describes the evolution of regolith developed by simple craters. Our results indicate that the regolith deposited on crater interiors is particularly important to consider when describing the distribution of regolith. Our model also indicates that the regolith thickness varies from one location to others. We apply this model to the regolith at the Apollo 15 landing site by considering the size distribution of observed small, simple lunar craters. Allowing for some regolith coming from outside of the area of the landing site, our result is consistent with an empirical result from the Apollo 15 seismic experiment.

## 1 Introduction

Since *Shoemaker et al.* [1967] first used the term 'regolith' to describe fragmented, low-cohesive units on the lunar surface, the formation of regolith has widely been recognized as a fundamental surface process on terrestrial bodies. On the Moon, while volcanic ash and beads [*Heiken et al.*, 1991] and thermal fatigue processes [*Delbo et al.*, 2014; *Molaro et al.*, 2015] may contribute to regolith growth, impact cratering has likely been the primary mechanism for regolith growth [*Shoemaker et al.*, 1969].

Observational studies have provided estimates of the thickness of lunar regolith. Using Lunar Orbiter 1 and Surveyor 1 photographs, *Oberbeck and Quaide* [1967] and *Quaide and Oberbeck* [1968] analyzed the crater morphologies and determined the regolith thickness on the Moon as 5–10 m. Active seismic experiments during the Apollo missions 11, 12, 14, 15, and 16 showed that the regolith thickness at the landing sites was on the order of meters [*Nakamura et al.*, 1975]. With 70-cm Arecibo radar data, *Fa and Wieczorek* [2012] measured the dielectric constant of the lunar surface and obtained that the median value of the regolith thickness ranged from 2.6 to 12.0 m, depending on geological terrain. *Wilcox et al.* [2005] estimated a regolith thickness of 8–30 m by analyzing Lunar Orbiter high-resolution images and Clementine multispectral images. Also, studies using Lunar Reconnaissance Orbiter Camera images revealed that the median regolith thickness in mare regions (~ 2 to 3 m) is shallower than that in highland regions (~ 5 to 8 m) [*Bart et al.,* 2011; *Fa et al.*, 2014]. Furthermore, lunar penetrating radar observations by the Chang'E mission revealed a layered structure at the rim of Chang'E crater located in Mare Imbrium: a reworked zone (< 1 m), an ejecta layer (~2 – 6 m), and a paleoregolith layer (~ 4 – 11 m) [*Fa et al.*, 2015].

Detailed modeling can provide a better understanding of regolith formation due to impact cratering. It is necessary to consider how the formation process of regolith is related to impact cratering processes and how cratering breaks down bedrock into regolith. When a small, simple crater forms into hard rock, regolith is produced by fragmentation of the target, and regolith is deposited in the crater's ejecta blanket and within its transient crater cavity [*Shoemaker et al.*, 1967]. The regolith unit within the crater cavity, i.e., the breccia lens, is developed during the excavation stage, during which time a bowl-shaped transient crater gravitationally collapses, followed by a process that loose materials move down on the steep wall of the transient crater [*Melosh*, 1989] and change in volume via dilatancy [*Collins*, 2014].

The condition of a cratered surface is usually compared to analyze the formation of regolith. We introduce a concept for a cratered surface that is heavily bombarded. As an example, we consider a hypothetical planetary surface that is initially intact. When this surface is constantly affected by impact cratering events, all the craters are initially visible. However, as the number of craters increases on the surface, some craters start to be degraded. Eventually, some crater sizes reach a situation that the total number of erased craters is equal to that of newly created craters. We call this condition crater equilibrium [*Melosh*, 1989].

Studies attempted to quantify regolith growth due to multiple impact events. *Shoemaker et al.* [1969] considered crater equilibrium to roughly estimate the thickness of regolith. Monte Carlo analyses have been a popular tool for simulating the evolution of regolith. Considering that the amount of regolith might change due to the condition of a target surface [*Quaide and Oberbeck*, 1968], *Oberbeck et al.* [1973] developed a Monte Carlo model to analyze how regolith grows in response to impact cratering. We note that while many sophisticated Monte Carlo techniques have recently been developed for different purposes, the Cratered Terrain Evolution Model (CTEM) is another Monte Carlo code that has provided strong insights into the regolith evolution on the Moon [*Richardson*, 2009; *Minton et al.*, 2015; *Huang et al.*, 2017].

The purpose of this paper is to characterize the evolution of the regolith thickness analytically. Numerical Monte Carlo techniques allow for describing complex regolith generation processes but tend to become computationally expensive. Furthermore, many of them still include imperfect functions of how a cratered surface on an airless body evolves because additions of complex physics usually induce redundant model uncertainties. For this reason, analytical models are useful for exploring the fundamental mechanisms in a simple way. Importantly, the thicker the regolith thickness grows, the more difficult it is for new small, simple craters to generate new regolith. Eventually, craters with any sizes no longer contribute to the regolith evolution. In this paper, we compute the time evolution of the regolith thickness and compare it with the time evolution of crater equilibrium.

We describe how we organize this paper. We first explain how to describe regolith units in a small, simple crater and how to compute the regolith thickness due to the formation of multiple craters. Next, we conduct comparison tests of our analytical model with the Monte Carlo simulation technique by *Oberbeck et al.* [1973] and with CTEM. Then, we apply this model to regolith formation at the Apollo 15 landing site. In this application, we consider that the crater size contributing to regolith generation by the formation of a breccia lens is different from that by ejecta blanketing. Some ejecta may be originated from large craters that are located far from the Apollo 15 site, while the formation of a breccia lens always results from a local process. Based on our crater counting, we consider that craters smaller than 100 m in radius affected regolith formation due to regolith infilling on the transient crater cavity, while those smaller than 381 m in radius did so by ejecta blanketing. For simplicity, however, we do not specify the minimum crater radius and simply assume that it is infinitely small, i.e., zero. Finally, using this application, we evaluate the timescale that the regolith thickness ceases to grow.

## 2 Modeling of regolith units as a source of regolith

### 2.1 Implementation of regolith units in a small, simple crater

To account for regolith formation due to the formation of small, simple craters, we track regolith production by these impacts within the ejecta blanket and breccia lens (Figure 1). Newly formed craters are assumed to be circular. For simplicity, we do not distinguish the size distribution of regolith particles in between a breccia lens and that an ejecta blanket. This simplification ignores the condition of regolith such as the size to reduce model uncertainties. Below, we introduce the regolith thickness profiles in the regions outside and inside the crater. In the outside region, we consider an ejecta blanket to be the regolith unit. In the inside region, a breccia lens is a source of regolith.

First, we discuss an ejecta blanket as a source of regolith. Earlier works showed the scaling law of the external crater profile as a function of distance from the crater rim [*McGetchin et al.*, 1973; *Pike*, 1974; *Fassett et al.*, 2011; *Xie and Zhu*, 2016]. We define the thickness profile of the ejecta blanket of a simple crater as [e.g., *Pike*, 1974; *Sharpton*, 2014]

$$h_{out} \sim \sigma r_c \left(\frac{r}{r_c}\right)^{-\kappa}, \qquad r > r_c, \tag{1}$$

where $r_c$ is the final crater radius, $r$ is the distance from the crater center, and $\sigma$ and $\kappa$ are empirical parameters. An observed height of an ejecta blanket may include the height of rim uplift and the thickness of ejecta. Therefore, it is necessary to find the part of the ejecta blanket correctly. In literature, the ejecta thickness at the rim is usually considered to be a half of the total rim height [*Melosh*, 1989]. However, recent works using high-resolution images revealed that for lunar craters and martian complex craters, the ejecta deposit, which is considered to be a part of the regolith unit in this study, might be much thinner than thought [*Sharpton*, 2014; *Sturm et al.*, 2016]. Particularly, *Sharpton* [2014] described a rim height of 0.068 $r_c^{1.01}$ and argued that the thickness of the ejecta blanket of a simple crater can be given as 0.014 $r_c^{1.01}$ $(r/r_c)^{-3}$. We use his scaling relationship but assume $r_c^{1.01} \sim r_c$ for mathematical simplicity; $\sigma$ and $\kappa$ are given as 0.014 and 3. Note that in this study, since the considered crater radius affecting regolith production due to ejecta blanketing is up to 381 m (see the discussion below), the assumption of $r_c^{1.01} \sim r_c$ only changes less than 6% of the thickness of each ejecta blanket, which is considered to be small.

Second, we add a model for creation and deposition of a breccia lens within a small, simple crater. A breccia lens forms in the area above the transient crater surface and beneath the final crater surface. We define the depth of the regolith unit within the breccia lens, using a parabolic profile, which is given as

$$h_{in} = \begin{cases} \delta\, r_c \left(1 - \dfrac{r^2}{r_c^2}\right), & r \le r_c \\ 0, & r > r_c \end{cases}, \tag{2}$$

where $\delta$ is the depth parameter. Recent works using Lunar Reconnaissance Orbiter Camera (LROC) images suggest that fresh small, simple craters have a low depth-to-diameter ratio [*Stopar et al.*, 2017; *Mahanti et al.*, 2018]. For example, the depth-to-diameter ratio is about 0.13 for fresh simple craters of 20 m – 50 m in radius and approximately 0.15 for those of 50 m – 100 m in radius [*Stopar et al.*, 2017]. The transient crater's depth is 1/3 of its diameter, and the final crater's diameter is 1.25 times larger than the transient crater's diameter. Using these conditions, we calculate $\delta$ as 0.34 for the 0.13 depth-to-diameter ratio and that as 0.30 for the 0.15 depth-to-diameter ratio.

## 2.2 Analytical expressions of regolith formation

In this section, we derive an analytical expression to describe the production of regolith from small, simple craters. Regolith units developed by such craters are described in Equations (1) and (2). A key factor in our modeling is to consider the regolith distribution of small, simple craters, accounting for the crater size distributions [*Costello et al.*, 2017; *Hirabayashi et al.*, 2017]. We introduce assumptions made in the present model. First, impacts randomly occur on an intact surface. Second, the target surface is assumed to be initially flat. Third, each small, simple crater only creates regolith units described by the profiles in Equations (1) and (2). Fourth, based on the third assumption, the target surface is not affected by other processes such as material melting or compaction.

We start our derivation by describing the fraction of a flat area covered by small, simple craters in some region at a given depth, $h$. We extend the work done by *Gault et al.* [1974], who considered this area fraction on the surface of the target region. Because craters have variable radii, we denote the radius of the $i$-sized crater as $r_{ci}$ $(i_{min} \leq i \leq i_{max})$, where $i_{min}$ and $i_{max}$ are the indices of the minimum crater radius and the maximum crater radius, respectively. We now consider that the target region is affected by craters of a single size $i$. The area fraction at $h$ that is affected by one crater of size $i$ is given as

$$P_i(h, r_{ci}) = \frac{S_i(h, r_{ci})}{A}, \tag{3}$$

where $S_i$ is the area at depth $h$ affected by a crater of radius $r_{ci}$, and $A$ is the total area of the target region. $S_i$ is a function of $h$ and $r_{ci}$, and so is $P_i$. We omit $(h, r_{ci})$ to simplify the mathematical expressions in the following process. We obtain the area fraction that is not occupied by a single crater of size $i$ as [*Gault et al.*, 1974]

$$\bar{P}_i = 1 - P_i = 1 - \frac{S_i}{A}. \tag{4}$$

Allowing for a number, $n_i$, of randomly emplaced craters of size $i$ in the same area $A$, we give the area fraction that is not occupied by any crater of this size as

$$\bar{P}_i^{n_i} = \left(1 - \frac{S_i}{A}\right)^{n_i}, \tag{5}$$

which for large $n_i$ is equivalent to

$$\bar{P}_i^{n_i} = \exp\left(-\frac{S_i n_i}{A}\right). \tag{6}$$

To extend this empty area fraction to all crater sizes, we form the product of $\bar{P}_i^{n_i}$ for all $i$s and rewrite as a summation using the law of indices, which is given as

$$P_e = \prod_{i=i_{min}}^{i_{max}} \bar{P}_i^{n_i} = \prod_{i=i_{min}}^{i_{max}} \exp\left(-\frac{S_i n_i}{A}\right) = \exp\left(-\frac{1}{A}\sum_{i=i_{min}}^{i_{max}} S_i n_i\right). \tag{7}$$

Hence, we can express the area fraction that is occupied by at least one crater of any size, $P = 1 - P_e$, as

$$P = 1 - \exp\left(-\frac{1}{A}\sum_{i=i_{min}}^{i_{max}} S_i n_i\right). \tag{8}$$

This quantity describes the area affected by any craters that have formed. Some new craters may occur in empty regions, while others may be in the areas that have already been affected by previously formed craters. This equation accounts for such crater overlapping (Figure 2).

We then convert Equation (8) to a continuous form. To do so, we describe $n_i$ as [*Hirabayashi et al.*, 2017]

$$n_i = -\frac{dC_t}{dr_c} dr_c, \tag{9}$$

where $C_t$ is the cumulative size frequency distribution (CSFD), which is given as [*Hirabayashi et al.*, 2017]

$$C_t = A\xi X r_c^{-\eta}, \tag{10}$$

where $r_c$ has units of m (i.e., meters), $\xi$, of which units are $\mathrm{m}^{\eta-2}$, is a factor that describes the produced crater CSFD of craters with a radius of 1 m per area at present, $\eta$ is the power index of the slope, and $X$ is the non-dimensional produced crater number based on a constant impact flux. From these definitions, the produced crater CSFD is proportional to $X$. $C_t$ at $X = 1$ indicates the produced crater CSFD at present.

We denote a continuous form of $S_i$ as $S$. We also define the region affected by an ejecta blanket as $S_{out}$ and that by a breccia lens as $S_{in}$. With these denotations, $S$ at given depth, $h$, is given as $S_{out} + S_{in}$ . For $S_{out}$ at $h$, we use Equation (1) to obtain

$$S_{out} = \pi r_c^2 \left(\frac{h}{\sigma r_c}\right)^{-\frac{2}{\kappa}} - \pi r_c^2. \tag{11}$$

In this form, we subtract $\pi r_c^2$ to remove the internal area of the crater. For $S_{in}$ at $h$, we use Equation (2) to obtain

$$S_{in} = \pi r_c^2 \left(1 - \frac{h}{\delta r_c}\right). \tag{12}$$

Under these conditions, we rewrite Equation (8) as

$$P = 1 - \exp\left\{\frac{1}{A}\int_{\frac{h}{\delta}}^{r_{max}^{in}} \frac{dC_t}{dr_c} S_{in} dr_c + \frac{1}{A}\int_{\frac{h}{\sigma}}^{r_{max}^{out}} \frac{dC_t}{dr_c} S_{out} dr_c\right\}, \tag{13}$$

where the minimum crater radius is characterized by $h$. The minimum radius of the breccia lens at $h$ is $h/\delta$, and that of the ejecta blanket is given as $h/\sigma$. For simplicity, we assume $h$ to range from zero to its maximum value; thus, this model accounts for infinitely small, simple craters, i.e., $r_c \to 0$ m. $r_{max}^{in}$ is the maximum radius of craters whose breccia lens affects the regolith thickness, while $r_{max}^{out}$ is that of craters whose ejecta blanket contributes to regolith generation. We introduce these parameters because the crater size contributing to regolith production by regolith infilling on the transient crater cavity is different from that due to ejecta blanketing. We use Figure 3 to explain this case. Assume that there are two craters formed on a test area. The large crater is younger than the small crater. The light gray areas describe ejecta blankets, while the dark gray regions indicate breccia lenses. Now we want to compute the regolith thickness at $Pnt$ in the breccia lens region of the small crater. This location is also affected by the ejecta blanket of the large crater but not by the breccia lens of that crater. Therefore, to properly compute the regolith thickness, we should consider the ejecta blanket of the large craters and the breccia lens of the small crater. Extending this argument to a case of multiple craters, we find that $r_{max}^{in}$ is dependent on the distribution of craters at a selected point, while $r_{max}^{out}$ should be large because ejecta from large craters can reach there.

Because of these expressions, $P$ is finally described as a function of $h$. From Equations (10), (11), and (12), we obtain the exponential terms in Equation (13) as

$$\begin{aligned} f_{in} &= \frac{1}{A}\int_{\frac{h}{\delta}}^{r_{max}^{in}} \frac{dC_t}{dr_c} S_{in} dr_c \\ &= \frac{\xi X \eta \pi}{\eta - 2}\left\{(r_{max}^{in})^{-\eta+2} - \left(\frac{h_{in}}{\delta}\right)^{-\eta+2}\right\} \\ &\quad - \frac{\xi X \eta \pi}{\eta - 1}\left(\frac{h_{in}}{\delta}\right)\left\{(r_{max}^{in})^{-\eta+1} - \left(\frac{h_{in}}{\delta}\right)^{-\eta+1}\right\}, \end{aligned} \tag{14}$$

$$\begin{aligned} f_{out} &= \frac{1}{A}\int_{\frac{h}{\sigma}}^{r_{max}^{out}} \frac{dC_t}{dr_c} S_{out} dr_c \\ &= -\frac{\xi X \eta \pi}{\eta - 2}\left\{(r_{max}^{out})^{-\eta+2} - \left(\frac{h_{out}}{\sigma}\right)^{-\eta+2}\right\} \\ &\quad + \frac{\xi X \eta \pi}{\eta - 2 - \frac{2}{\kappa}}\left(\frac{h_{out}}{\sigma}\right)^{-\frac{2}{\kappa}}\left\{(r_{max}^{out})^{-\eta+2+\frac{2}{\kappa}} - \left(\frac{h_{out}}{\sigma}\right)^{-\eta+2+\frac{2}{\kappa}}\right\} \end{aligned} \tag{15}$$

where

$$h_{in} = \min(h, \delta r_{max}^{in}), \tag{16}$$

$$h_{out} = \min(h, \sigma r_{max}^{out}). \tag{17}$$

Equation (13) describes the fraction of the area affected by at least one crater formation at $h$ to the total area. This area fraction indicates how much regolith is filled in the total area at $h$. It also describes the total area whose thickness is thicker than $h$. Using this area fraction to express the regolith thickness distribution, we can directly compare the results from our model with those from *Oberbeck et al.* [1973], who used this quantity (e.g., see Figures 1 and 3 in their paper).

While Equation (13) indicates the distribution of the regolith thickness, this equation does not give an explicit insight into the regolith thicknesses at sampling locations that would be restricted by many artificial factors such as safety for astronauts and accessibility. Instead, we can consider the expected regolith thickness at the sample locations in a statistical sense. We consider sampling locations randomly chosen from the region of the landing site. This process may be described by the expected value of a random variable in probability theory, which is given as

$$\begin{aligned} \bar{h} &= \lim_{\Delta h \to 0} \sum_{i=i_{min}}^{i_{max}} \left\{ P\left(h_i - \frac{\Delta h}{2}\right) - P\left(h_i + \frac{\Delta h}{2}\right) \right\} h_i , \\ &= -\int_{h(r_{min})}^{h(r_{max})} \frac{dP}{dh} h dh, \end{aligned} \tag{18}$$

where $\Delta h$ is a small interval of the depth, $i_{min}$ and $i_{max}$ are the indices, and $h(r_{min})$ and $h(r_{max})$ define the regolith thicknesses created by craters with radii of $r_{min}$ and $r_{max}$, respectively. Using the assumption above, we consider $r_{min}$ to be zero. $r_{max}$ can be replaced with either $r_{max}^{in}$ or $r_{max}^{out}$. To distinguish the regolith thickness value given in Equation (18) with the regolith thickness distribution, we call it the expected regolith thickness.

## 3 Benchmark exercises of the analytical model

### 3.1 Comparison test with the Monte Carlo analysis done by *Oberbeck et al.* [1973]

In this section, we compare the developed analytical model with the Monte Carlo simulation work done by *Oberbeck et al.* [1973]. *Oberbeck and Quaide* [1967] and *Quaide and Oberbeck* [1968] analyzed the crater topographic profiles to determine the regolith thickness, using observational and experimental data. *Oberbeck et al.* [1973] applied the analytical expressions of the regolith volume produced by crater formation [*Quaide and Oberbeck*, 1968] to explore the evolution of the regolith thickness. In their work, the total regolith volume was converted to the ejecta profile function, which corresponds to Equation (1). Here, we strictly follow their simulation settings. *Oberbeck et al.* [1973] considered a case with a produced crater CSFD of $1.0 \times 10^7 D^{-3.4}$ per km$^2$ (Figure 1 in their paper), where $D$ is the crater diameter in their definition. From this function, $\xi$ and $\eta$ are defined as 0.95 and 3.4, respectively.

We first consider the coefficient of the regolith thickness that they considered, which is given as $C$ in their work (see Equation (2) in *Oberbeck et al.* [1973]) and as $\sigma$ in the present work.

To characterize how the regolith thickness of the pre-impact surface affects the newly generated regolith volume, *Oberbeck et al.* [1973] defined four categories: (I) regolith crater, (II) a flat-bottomed crater, (III) a concentric crater, and (IV) a hard-rock crater. We compute the regolith volume of these groups, using Table 1 in *Oberbeck et al.* [1973] (see Table S1 in our supporting information). Then, we calculate $\sigma$ for *Oberbeck et al.* [1973], which is denoted as $\sigma_{ob}$, by using Equation (4) in *Oberbeck et al.* [1973]:

$$\sigma_{ob} = \frac{V(2-\kappa)}{2\pi r_c^3 (q^{2-\kappa} - 1)}. \quad (19)$$

where $V$ is the regolith volume defined in Table 1 in *Oberbeck et al.* [1973], $q = 7$, and $\kappa = 3.7$. Note that $\kappa$ used in Equation (19) is equal to that in Equation (1). To compute $V$, we follow the parameters defined by *Oberbeck et al.* [1973] (Table S2). We find that $\sigma_{ob}$ is not strongly affected by the crater size and the crater floor size. Considering $\sigma_{ob}$ for all the cases (Table S3), we obtain the averaged value of $\sigma_{ob}$ as 0.15 (Text S1).

Again, given $\kappa$ as 3 and the rim height-to-radius ratio as 0.068, *Sharpton* [2014] proposed $\sigma = 0.014$, while $\sigma = 0.034$ in literature, which is the half of the rim height [e.g., *Melosh*, 1989]. Because *Oberbeck et al.* [1973] defined $\kappa$ as 3.7, we need to fix the discrepancy of $\kappa$ to compare these $\sigma$s with $\sigma_{ob}$. Based on Equation (19), a formula to convert $\sigma$ from one slope case to another is given as

$$\sigma^* = \frac{(2-\kappa^*)}{(2-\kappa)} \frac{(q^{2-\kappa} - 1)}{(q^{2-\kappa^*} - 1)} \sigma, \quad (20)$$

where superscript $*$ gives the post-conversion case. Here, we convert $\sigma$ from $\kappa = 3$ to $\kappa^* = 3.7$. We denote $\sigma = 0.014$ as Case 1 and $\sigma = 0.034$ as Case 2. Using Equation (20), we obtain $\sigma^* = 0.021$ for Case 1 and $\sigma^* = 0.051$ for Case 2 (Table 1). For comparison, Table 1 also describes the ratios of the rim height to the crater radius. It is found that since $\sigma_{ob} = 0.15$, $\sigma^* = 0.021$ for Case 1 is 7.1 times smaller than $\sigma_{ob}$, and $\sigma^* = 0.051$ for Case 2 is 2.9 times smaller than $\sigma_{ob}$. This result shows that *Oberbeck et al.* [1973] overestimated the ejecta blanket volume.

Accounting for an ejecta blanket and a breccia lens as sources of regolith, we compare our analytical model with their Monte Carlo simulation. The defined parameters used for the analytical model are given in Table S4. Figure 4 shows the area fractions obtained by comparison between our analytical results and the Monte Carlo simulation results from *Oberbeck et al.* [1973]. The red and blue lines describe the results from our analytical model with different $\sigma^*$s. The blue line is $\sigma^* = 0.021$, while the red line is $\sigma^* = 0.051$. We find that a difference between these $\sigma^*$s is negligibly small and conclude that an ejecta blanket plays a minor role in producing regolith.

On the other hand, the black dashed lines with gray-colored squares show the results from *Oberbeck et al.* [1973]. The lower black dashed line is the case when the minimum crater size is 94.8 m, and the upper one is the case when the minimum size is 9 m. This variation resulted from a resolution issue in their Monte Carlo model and was improved by considering a higher resolution [*Oberbeck et al.*,1973]. Thus, because the analytical model accounts for an infinitely small crater size, our results are expected to become closer to the upper black dashed line. However, they are lower at a high area fraction and higher at a low area fraction than the upper black dashed line. This discrepancy results from the sources of regolith generation; while *Oberbeck et al.* [1973] only considered the ejecta volume, the present work accounts for the volume of an ejecta blanket and a breccia lens. This discrepancy is likely to change the distribution of regolith at given depth. Importantly, the expected regolith thickness of our model is 2.0 m, which corresponds to that of the upper dashed line obtained by *Oberbeck et al.* [1973]. While *Oberbeck and Quaide* [1967] and *Quaide and Oberbeck* [1968] analyzed the crater profiles to determine the regolith thickness, *Oberbeck et al.* [1973] assumed that regolith production was solely attributed to ejecta blanketing. We infer that *Oberbeck et al.* [1973] might overestimate the ejecta volume.

3.2 Comparison tests with Cratered Terrain Evolution Model

Next, we compare the results from our analytical model with those from CTEM. We first describe the implementation of the profile of a breccia lens into CTEM. The original version of CTEM has a function that creates an ejecta blanket based on the ejecta volume and speed and the size of each newly formed crater [*Richardson*, 2009; *Minton et al.*, 2015]. We use this function in this work to compute the ejecta blanket profile. We note that while *Huang et al.* [2017] incorporated crater ray patterns into CTEM to model distal ejecta, we treat ejecta emplacement to be uniform in our analytical model. We implement a new function that calculates the breccia lens profile (see Equation (2)), using a linked list algorithm first incorporated by *Huang et al.* [2017]. Since CTEM has an algorithm that computes a crater shape profile [*Richardson*, 2009; *Minton et al.*, 2015], we merged a new shape profile of the transient crater into this algorithm. The ratio of the transient crater depth to the transient crater diameter is defined as 1/3. Also, we set the depth-to-diameter ratio in CTEM as 0.13, which leads to $\delta = 0.34$.

Figure 5 shows the regolith thickness after formation of a small, simple crater with a radius of 38.2 m. Figure 5a is the case when only an ejecta blanket is considered to be a source of regolith, while Figure 5b is the case when both an ejecta blanket and a breccia lens result in the formation of regolith. Equation (1) predicts an ejecta thickness of 0.53 m at the rim, while CTEM produces that of 0.48 m there (Figure 5a). This difference results from CTEM's ejecta generation function, which is not easily controllable. Because the error is 10 % of the ejecta thickness, we keep the original CTEM setting. Also, the regolith thickness pattern in Figure 5a comes from CTEM's periodic boundary condition [*Richardson*, 2009; *Minton et al.*, 2015]. On the other hand, the regolith thickness in the breccia lens becomes 13 m at the crater center from Equation (2), and we produce the same regolith thickness in CTEM (Figure 5b). In CTEM, the regolith layer profiles are stored in the linked list and are updated when a new crater affects the originally existing bedrock layer. This function allows CTEM to compute realistic regolith generation.

CTEM generates craters using multiple functions including a model impactor velocity distribution, a model impactor size distribution, and a crater scaling relationship. As a result of a scheme based on these functions, the produced crater CSFD is consistent with a lunar-based produced CSFD [*Minton et al.*, 2015]. A surface area defined in CTEM is defined by 500 m by 500 m. In each simulation, craters are generated in the radius range between 2 m and 100 m. We compute 50 statistical cases with different random seeds. While the produced crater CSFD in each case is different, the mean produced crater CSFD of this statistical test is given using Equation (10) with $A = 25 \times 10^4$ m$^2$, $\xi = 0.80$ m, $\eta = 3.0$, and $X = 1.0$. CTEN generated nearly 9000 craters in each case.

Similar to the previous section, we use the area fraction to compare the results from our analytical model with those from CTEM (Figure 6). The gray lines indicate the results from CTEM, and the dashed black lines describe the means of these results. The solid black lines represent the results from the analytical model. The parameter setting of the analytical model is available in Table S5. Figure 6a is the produced crater CSFDs, and Figure 6b is the area fraction of regolith at given depth. The solution from the analytical model and the means of the results from CTEM are consistent. However, we note that in Figure 6b, the dashed black line shows a slightly smaller regolith area fraction at deep depth (> 3 m) than the solid black line. We discuss this discrepancy by considering the produced crater CSFD. As shown in Figure 6a, the produced crater CSFD for the analytical model (the solid black line) is higher than that for CTEM (the dashed black line) at large craters (> 30 m in radius). Because of this effect, the analytical model apparently predicts a slightly wider area fraction in a deep region than CTEM. We also observe a slight difference between our analytical model and CTEM in a shallow region. We infer that CTEM accounts for the topographic evolution of a cratered surface, but the analytical model does not. This model discrepancy gives an inconsistency of the area fraction at shallow depth in CTEM and the analytical model.

## 4 Applications of the analytical model to the Apollo 15 landing site

This section applies the analytical model to the regolith evolution at the Apollo 15 landing site. This region, located at a boundary between highlands and mare, is considered to have experienced various processes that contributed to its unique geology (Figure 7). The main unit at the landing site is mare lava emplaced 3.3 Ga ago [*Nyquist and Shih*, 1992]. Aristillus and Autolycus, complex craters that formed after local emplacement of mare basalt inside Imbrium basin, are located 300 km away from the landing site [*Bernatowicz and Hohenberg*, 1978; *Ryder et al.*, 1991]. This area is also surrounded by the Apennine Bench Formation, which is thought to have formed by either impact [*Wilhelms*, 1980; *Mccauley et al.*, 1981] or volcanic activity [*Spudis*, 1978; *Spudis and Ryder*, 1985]. Here, we first conduct crater counting of the landing site to obtain the visible crater CSFD (Section 4.1). Next, comparing the results from the analytical model with the empirical data obtained in Section 4.1, we investigate the regolith thickness evolution at the Apollo 15 landing site (Section 4.2). Then, we explore if the regolith thickness further grows at this site by considering the timescale of the regolith thickness evolution (Section 4.3).

### 4.1 Counting craters

We use the shaded relief map and the digital terrain model (DTM) for the Apollo 15 landing site (http://wms.lroc.asu.edu/lroc/view_rdr/NAC_DTM_APOLLO15), derived from LROC Narrow Angle Camera (NAC) images M111571816L/R and M11578606L/R [*Tran et al.*, 2010]. The DTM post spacing is 2.0 m. We consider a 4.7 km × 10.6 km region in the northern portion of the map (Figure 8). The Falcon lander is located in a southern portion of the full maps, so the currently considered area does not directly include the Apollo 15 site. Because our technique explicitly requires the crater density, we must measure the surface area. For this reason, we choose the northern region to avoid complex surface features such as Hadley Rille and Montes Apenninus (see Figure 7). Note that the crater conditions there, such as the size frequency distribution and freshness, are similar to those in the region around the Apollo 15 site.

To count craters, we use publicly available remote sensing software, JMARS [*Christensen et al.*, 2009]. To find craters, we search for their rim feature and circular depression. The shaded relief and DTM maps enable us to identify topographic depression developed by impact cratering. Note that in our crater counting process, primary and secondary craters are not distinguished. Because some secondary craters may be morphologically indistinguishable from small primary craters [*Wells et al.*, 2010], the counted craters presumably include some of these craters. We count craters with radii larger than 3.8 m (the empirical data is available in our supporting information). Figure 9 shows the visible crater CSFD of craters larger than 12 m in radius to avoid crater counting bias. The red-edged circles in this figure show our empirical result. We interpret the observed different slopes as two crater conditions. The steep slope is indicative of the produced crater CSFD, while the shallow slope describes crater equilibrium. Fitting the crater equilibrium model by *Hirabayashi et al.* [2017] to our empirical result, we determine the produced crater CSFD given in Equation (10) and the equilibrium slope, $C_c^\infty$, as

$$C_t = 1.1 \times 10^8 \, r_c^{-3.2}, \tag{21}$$

$$C_c^\infty = 1.2 \times 10^5 \, r_c^{-1.5}, \tag{22}$$

respectively. Similar to Equation (10), the units of $r_c$ of these equations are meters. Since we consider that $X = 1$ and $A = 49.8$ km$^2$, we obtain $\xi = 2.21$ m$^{1.2}$. Therefore, the produced crater CSFD per area is given as $2.21 \, r_c^{-3.2}$ m$^{-2}$. These two fitting functions indicate that in this region, craters with radii less than 50 m are in crater equilibrium. The obtained results are consistent with the crater counting analysis by *Robbins et al.* [2014] and the crater equilibrium study by *Hirabayashi et al.* [2017].

## 4.2 Regolith thickness at present

Impact craters have dominated the landscape of the Apollo 15 landing site since the total reset by a mare lava flow, suggesting that they are the main control on the regolith thickness in the landing site region. As a consequence, the regolith thickness is an indicator of impact cratering events after this region was flooded with mare lava 3.3 Ga ago [*Nyquist and Shih*, 1992]. We compute the evolution of the regolith thickness at this landing site. We assume that the produced crater CSFD given in Equation (21) represents the local region around the lunar module. From a passive seismic experiment in the Apollo 15 mission, *Nakamura et al.* [1975] obtained the regolith thickness as 4.4 m.

Before discussing our results, we introduce the values of $r_{max}^{in}$ and $r_{max}^{out}$. We first determine $r_{max}^{in}$. We choose an area within 100 m from the lunar module landing site to recover the passive seismic experiment distance done by the Apollo 15 crews [*Nakamura et al.*, 1975]. This area is surrounded by degraded craters of ~100 m in radius. Here, we consider that craters similar to or smaller than these craters might affect regolith formation due to regolith infilling on the transient crater cavity. Thus, we define $r_{max}^{in}$ as 100 m. Second, for $r_{max}^{out}$, it is necessary to account for the distribution of large craters in a wide area around the landing site. In our counting, the largest crater is 381 m in radius. Therefore, to consider the effect of ejecta from all the observed craters, we define $r_{max}^{out}$ as 381 m.

We note that there are several large craters whose ejecta might affect the Apollo 15 landing site. While our model can account for ejecta from large craters by choosing a proper value of $r_{max}^{out}$ for their crater radii, we consider that our produced crater CSFD may not represent the global distribution of large craters. Here, we assume that $C_t$ in Equation (21) does not account for the large crater distribution; therefore, we separate the effect of these craters on regolith growth at the landing site from our analytical model. St. George crater, which is the largest crater (1000 m in radius) in the local area, is located 5 km from the landing site. Using the ejecta thickness relationship for a simple crater given by *Sharpton* [2014], we obtain the ejecta thicknesses from this crater as 0.12 m. Autolycus and Aristillus craters are located 177 km and 275 km from the landing site, respectively [*Smith et al.*, 2010]. The radius of Autolycus is $\sim 16.5$ km, while that of Aristillus is $\sim 27.5$ km. Applying the ejecta thickness relationship for a complex crater given by *Sharpton* [2014], we obtain the ejecta thickness from Autolycus and Aristillus as 0.15 m and 0.23 m, respectively. If this is the case, 11% of the regolith thickness at the landing site, i.e., ~0.5 m, might result from these craters. The $FeO$ content derived from Clementine imagery implies that the materials at the landing site might have been affected by exotic materials, but the influence was still limited [*Lucey et al.*, 1995; *Blewett et al.*, 1997].

Figure 10 shows the area fraction derived from the analytical model with $\delta = 0.32 \pm 0.02$. The red line is the case when only an ejecta blanket is considered, while the black line with a gray area is the case when both an ejecta blanket and a breccia lens affect regolith formation. The results are obtained using $\delta$ that is between 0.30 and 0.34; the upper bound of the gray area is at $\delta = 0.34$, the lower bound is at $\delta = 0.30$, and the black line in the middle gives the case of $\delta = 0.32$. For the ejecta-only case, because the result is independent of $\delta$, it is uniquely determined. If an ejecta blanket is the only source of regolith, the regolith thickness is too shallow. On the other hand, if regolith is generated by both an ejecta blanket and a breccia lens, the amount of regolith significantly increases. We compute the expected regolith thickness for the ejecta-only case as 0.41 m and that for the ejecta-breccia lens case as $3.85 \pm 0.25$ m without ejecta coming from the outside of the landing site. This result indicates that ejecta blanketing does not contribute to regolith generation. Thus, given an addition of some exotic ejecta (~0.5 m), the case accounting for both an ejecta blanket and a breccia lens is consistent with the empirical study by *Nakamura et al.* [1975]. The parameters for the analytical model are described in Table 2.

Finally, we argue that it is difficult to infer the regolith thickness from the crater equilibrium condition. *Shoemaker et al.* [1969] computed an upper limit of the regolith thickness by looking at the largest crater size that reaches crater equilibrium. Any craters smaller than this size have completely been lost due to crater degradation. Thus, they hypothesized that these craters should have experienced a regolith filling process on their final crater surfaces, and thus the excavation depth of the largest crater in crater equilibrium would be an upper bound of the regolith thickness [*Shoemaker et al.*, 1969]. As discussed in Section 4.1, craters smaller than 50 m in radius are likely to be in crater equilibrium at the Apollo 15 landing site. Since the depth-to-diameter ratio is 0.13 for simple craters with a radius between 20 m and 50 m [*Stopar et al.*, 2017], the excavation depth of the 50-m-radius crater is 13 m, and so is the regolith thickness. This regolith thickness is higher than the regolith thickness reported by *Nakamura et al.* [1975].

## 4.3 Regolith growth timescale

In this section, using our analytical model, we discuss if the regolith thickness still evolves at the Apollo 15 landing site. When a surface is initially intact, the regolith thickness rapidly increases. However, as the total number of craters increases, new craters are formed on old craters, and the amount of regolith generation gradually decreases. Eventually, the regolith thickness does not increase any longer. We define this evolution timescale as the regolith growth timescale. We describe the regolith thickness as functions of $X$ and the model age [*Neukum et al.*, 2001]. We first compute the expected regolith thickness as a function of $X$ by using Equation (18). Then, this thickness function is also expressed as a function of the model age. Here, we only consider the case of $\delta = 0.32$.

To quantify the timescale that regolith growth ends, we note that the size of craters controls the regolith growth timescale. As the model age increases, it is likely that large craters affect regolith growth at the landing site, and the regolith thickness grows further. Therefore, the timescale that regolith growth ceases depends on $r_{max}^{in}$ and $r_{max}^{out}$. Here, for simplicity, we show the regolith thickness evolution under an assumption that $r_{max}^{in}$ and $r_{max}^{out}$ are the same and constant over a considered timescale. This assumption may be unrealistic because these quantities may be time-variant over a long timescale; however, with this assumption, we can quantitatively show how the regolith growth depends on the maximum crater size.

We consider three different cases. The first case is $r_{max}^{in} = r_{max}^{out} = 100$ m, the second case is $r_{max}^{in} = r_{max}^{out} = 500$ m, and the third case is $r_{max}^{in} = r_{max}^{out} = 1000$ m. Figure 11 shows the time evolutions of the expected regolith thickness as a function of $X$ $(0 \leq X \leq 590)$ (Figure 11a) and that as a function of the model age (≤ 4.5 Ga) (Figure 11b). A flat slope in each case indicates that regolith growth ceases. The results show that although the regolith growth timescale at $r_{max}^{in} = r_{max}^{out} = 100$ m is the shortest, the regolith thickness in this case still increases rapidly at present. The regolith growth timescale becomes long as the maximum radius considered increases. Therefore, if the Apollo 15 landing site is affected by further impact events, the regolith thickness should increase in all these cases.

We compare the regolith growth timescale with the timescale of crater equilibrium. As discussed in Section 4.1, craters with a radius smaller than 50 m are in crater equilibrium at present (Figure 9). On the other hand, any cases tested above show that the regolith thickness still grows rapidly at present. Thus, we infer that the regolith growth timescale is significantly longer than the timescale of crater equilibrium. Note that our consideration does not account for other regolith formation processes such as volcanism; however, processes besides impacts are probably not relevant at the Apollo 15 landing site, at least after the period of mare emplacement [*Nyquist and Shih*, 1992].

## 5 Discussion

### 5.1 Timescales of regolith growth and crater equilibrium

We predicted that the regolith growth timescale is substantially longer than the crater equilibrium timescale at the Apollo 15 landing site. We propose that this timescale difference is attributed to how small crater formation can affect these processes. On a cratered surface, the topographic features are constantly degraded due to impact and thermal processes. Small craters play significant roles in topographic diffusion [*Fassett and Thomson*, 2014; *Fassett et al.*, 2017], and micrometeorite impacts break down rock surfaces [*Hörz and Cintala*, 1997]. Significant temperature variations in planetary surfaces induce thermal cracking [*Molaro et al.*, 2015]. The crater equilibrium timescale strongly depends on these processes. On the other hand, the regolith thickness is mainly controlled by how impact events fragment subsurface layers. The breccia lens of a pre-existing crater can be preserved until a larger crater is generated in the same location. It is important to note that a shock propagation during an impact cratering process is likely to damage a wider area [*Collins et al.*, 2004], implying that such a damaged area may be an additional source of regolith. This effect may increase the regolith growth rate. However, again, to generate additional regolith, new craters should be large enough to fragment intact layers.

We address that by quantifying the regolith growth timescale, we may be able to estimate the actual impact flux from the regolith depth even if this surface is in crater equilibrium. It has long been argued that a surface that is in crater equilibrium is likely to lose the impact cratering history, and thus it is difficult to infer the surface condition from empirical data [e.g., *Hartmann*, 1984]. However, since the timescale of regolith growth is much longer than that of crater equilibrium, the regolith thickness provides useful information of the impact cratering history. Further investigations will shed light on the cratered surface evolution.

5.2 Dependence of the thickness of a breccia lens on the depth parameter

In the previous sections, we define the depth parameter, $\delta$, as 0.30 – 0.34 by applying the ratio of the final crater depth to the final crater diameter obtained by *Stopar et al.* [2017]. *Stopar et al.* [2017] and *Mahanti et al.* [2018] proposed that small, simple craters might be shallower than large, simple craters. Specifically, the depth-to-diameter ratio is 0.13 for 20-to-50-m-radius craters, 0.15 for 50-to-100-m-radius craters, and 0.17 for 100-to-200-m-radius craters [*Stopar et al.*, 2017]. However, terrestrial craters with a radius smaller than 250 m may have a unique V-shaped crater cavity; such craters are a so-called Odessa-type [*Shoemaker et al.*, 2005]. This report implies that Odessa-type craters may have a high depth-to-diameter ratio (~0.25) and thus a thin breccia lens [*Shoemaker et al.*, 2005].

The formation of a breccia lens results from the process that regolith at the rim slides down on the crater wall [*Melosh*, 1989; *Collins*, 2014]. For terrestrial craters, *Shoemaker et al.* [2005] implied that the thin breccia lens of Odessa-type craters might result from an inefficient regolith flow after the excavation phase. For lunar craters, to explain a low depth-to-diameter ratio of fresh small, simple craters, *Mahanti et al.* [2018] proposed a short-term infilling process possibly due to topographic diffusion after the crater formation. This feature was also observed by *Basilevsky et al.* [2014]. To compute the thickness of a breccia lens, we rule out a possibility that the depth of the transient crater of a small, simple crater may be shallower than that of a large, simple crater. The ratio of the transient crater depth to the transient crater diameter for Odessa crater is similar to that for Meteor crater [*Short*, 2006].

Because these works propose different thickness of a breccia lens, we discuss how the regolith thickness depends on $\delta$. Using the same simulation settings used in Table 2, we compute the regolith area fraction at present for three different $\delta$s: $\delta = 0.1$, 0.2, and 0.34. Note that we give the case discussed in Section 4.2 as the third case for comparison. Figure 12 indicates how the area fraction varies due to $\delta$. The expected regolith thicknesses for the cases of $\delta = 0.1$, 0.2, and 0.34 are 1.2 m, 2.4 m, and 4.1 m, respectively. It is found that the regolith thickness is sensitive to the value of $\delta$. Also, we show the ejecta-only case in red. The regolith thickness in this case is shorter than that in other cases. This result indicates that a breccia lens plays a primary role in regolith generation.

*Stopar et al.* [2017] argued a possibility that the shallow depth-to-diameter ratio of fresh small, simple craters resulted from the inclusion of secondary craters. While these craters have been reported to have an influence on the surface morphology of the Moon and other planetary objects [e.g., *McEwen et al.*, 2005; *Bierhaus et al.*, 2005; *McEwen and Bierhaus*, 2006; *Robinson et al.*, 2015], one interpretation of the results by *Speyerer et al.* [2016] is that generation of primary craters is far more frequent than that of secondary craters [*Stopar et al.*, 2017]. If this is true, the majority of craters investigated by them should be primary craters, although measuring the depth-to-diameter ratio only cannot identify if primary or secondary craters played a role in the shallow depth-to-diameter ratio [*Stopar et al.*, 2017]. It may be possible that the volume of the breccia lens is the same in both Odessa-type craters on the Earth and small, simple craters with shallow depth-to-diameter ratios on the Moon. Thus, a different scaling relationship may provide consistent regolith growth on both terrestrial and lunar surfaces. An attempt at a successful explanation of these works is beyond our scope, although further improvement of the present model and investigations will shed light on this issue.

### 5.3 Volumes of ejecta and a breccia lens

It is reasonable to consider that if the porosity is constant, the total volume of materials should not change before and after the crater formation process. However, our parameter setting does not explicitly satisfy this volume conservation condition and increases the volume up to 11% from the initial material volume during one crater formation. This inconsistency comes from the fact that we use crater profile parameters independently derived for the ejecta thickness [*Sharpton*, 2014] and the depth-to-diameter ratio [*Stopar et al.*, 2017]. This process produced an additional material volume.

Also, because of the use of the ejecta thickness condition by *Sharpton* [2014], the volume fraction of the ejecta over the total regolith generated during one crater formation becomes small. In the simulation with the parameters in Table 2, the volume of the ejecta is only 15% of the total regolith volume, while that of the breccia lens is 85 %. In a classical view, the volumes of the ejecta and the breccia lens are considered to be almost equal [*Melosh*, 1989]. In addition, small-scale experiments by *Stöffler et al.* [1975] implied that ejecta might dominate regolith generation during a crater formation in quartz sands although it may be difficult to identify the ejecta from the rim-uplifting feature.

We test if the volume inconsistency and the small ejecta-volume fraction could affect our simulation. In this test, we modify the total rim height so that the material volume does not change. When the rim height is $0.053r_c$ (as comparison, $0.068r_c$ from *Sharpton* [2014]), the material volume is constant before and after one crater is formed. Also, we assume that all the volume in this rim height consists of ejecta and contributes to regolith generation. In these conditions, we obtain $\delta = 0.31$ and $\sigma = 0.053$. The fraction of ejecta reaches 42% in the total regolith volume, and that of a breccia lens becomes 58%. Figure 13 shows this volume-constant cases and compares them with the cases that use the parameters in Table 2 (the original case). We find that these cases do not have a significant difference. For the ejecta-breccia lens cases, the expected regolith thicknesses for the volume-constant case is 3.91 m, while that for the original case is 3.85 m. For the ejecta-only cases, the expected regolith thicknesses are 1.26 m and 0.41 m for the volume-constant case and the original case, respectively. This test suggests that the formation of a breccia lens mainly controls the total amount of impact-generated regolith.

### 5.4 Future improvements of the analytical model

We discuss open questions for the present analytical model. First, our analytical model does not consider the surface topographic evolution, although its analytical expression has been proposed in an earlier work [*Rosenburg et al.*, 2015]. Second, secondary craters were likely to be included in our counting. Because of the impact speed and possible clustering, these craters might have contributed to regolith generation differently. The present study cannot distinguish the effects of primary and secondary craters, although we emphasize the importance of quantifying these effects. Third, our model did not distinguish regolith in an ejecta blanket with that in a breccia lens. However, these regions may have different size distributions of regolith materials. Fourth, the formation of multiple craters may further fragment surface materials, causing the size distribution evolution of regolith materials. This consideration may be related to the mechanism of regolith mixing [*Gault et al.*, 1974; *Arnold*, 1975; *Speyerer et al.*, 2016; *Costello et al.*, 2017]. Fifth, the ejecta thickness scaling law that we applied is known to be oversimplified. *Huang et al.* [2017] showed that distal ejecta would affect mixing process based on the patterns of crater rays. Thus, such spatial heterogeneity would control the magnitude of regolith mixing. Finally, we address that the discrepancy of the volume fractions of ejecta and a breccia lens is a critical issue to reconcile our model with the recent findings of small, simple craters on the lunar surface. Further investigations of all these questions are our future work.

## 6 Conclusion

We developed a new analytical model that accounted for the effect of a breccia lens and an ejecta blanket to compute the evolution of the regolith thickness via the formation of multiple small, simple impact craters (< 381 m in radius for ejecta blanketing and < 100 m in radius for regolith infilling on the transient crater cavity). This analytical model was statistically designed to be able to account for the decrease in the volume of newly produced regolith due to crater overlapping. This model was evaluated using two Monte Carlo simulation techniques; we confirmed that a breccia lens played a more critical role in producing regolith than an ejecta blanket. We applied this analytical model to the regolith evolution at the Apollo 15 site. Our model predicted that regolith with a thickness of 3.85 m might be generated locally. Given some amount of externally originated regolith (~0.5 m), our result was consistent with a seismic experiment conducted by the Apollo 15 crews, 4.4 m. The results successfully showed that the regolith thickness was controlled by the size distribution of craters and varied spatially due to the stochastic location of specific impacts. Also, we found that the regolith thickness has not reached a steady state but might still increase with additional impacts. Because this surface is in crater equilibrium at present, we concluded that the timescale of regolith growth is longer than that of crater equilibrium.

## Acknowledgments, Samples, and Data

M.H. J.M.S., and D.A.M. acknowledge support by the NASA LDAP program (NNH15ZDA001N-LDAP). M.H. also thanks for support by an assistant professor startup funding from Department of Aerospace Engineering at Auburn University. C.I.F. acknowledges support by the NASA LDAP program. The authors thank Dr. Gareth Collins and anonymous referees for useful comments that substantially improved our manuscript. The crater counting data and the parameters used for the analytical model are found in supporting information.

## Tables

**Table 1**. Values of the depth parameter, $\sigma$, for different conditions. $\kappa$ indicates the slope of the ejecta thickness.

| | Rim height | Ejecta thickness at the rim | |
|---|---|---|---|
| | | *Sharpton* [2014] | Literature [e.g., *Melosh*, 1989] |
| $\boldsymbol{\kappa = 3.0}$ | 0.068 | 0.014 | 0.034 |
| $\boldsymbol{\kappa^{*} = 3.7}$ | 0.10 | 0.021 | 0.051 |

**Table 2**. Model parameters for the analytical model used for calculations of regolith thickness at the Apollo 15 landing site. [-] in Value means that the value is non-dimensional.

| Parameter | Symbol | Value [Units] |
|---|---|---|
| **Ratio of the breccia lens thickness ratio at the crater center to the crater radius** | $\delta$ | $0.32 \pm 0.2$[-] |
| **Ratio of the Ejecta thickness ratio at the crater rim to the crater radius** | $\sigma$ | 0.014 [-] |
| **Slope of the ejecta thickness** | $\kappa$ | 3.0 [-] |
| **Max. crater radius affecting regolith formation by regolith infilling on the transient crater cavity** | $r_{max}^{in}$ | 100 [m] |
| **Max. crater radius affecting regolith formation due to ejecta blanketing** | $r_{max}^{out}$ | 381 [m] |
| **Slope of the produced crater CSFD** | $\eta$ | 3.2 [-] |
| **Produced crater CSFD coefficient** | $\xi$ | 2.21 [$m^{1.2}$] |
| **Normalized crater number** | $X$ | 1.0 [-] |

**Figures**

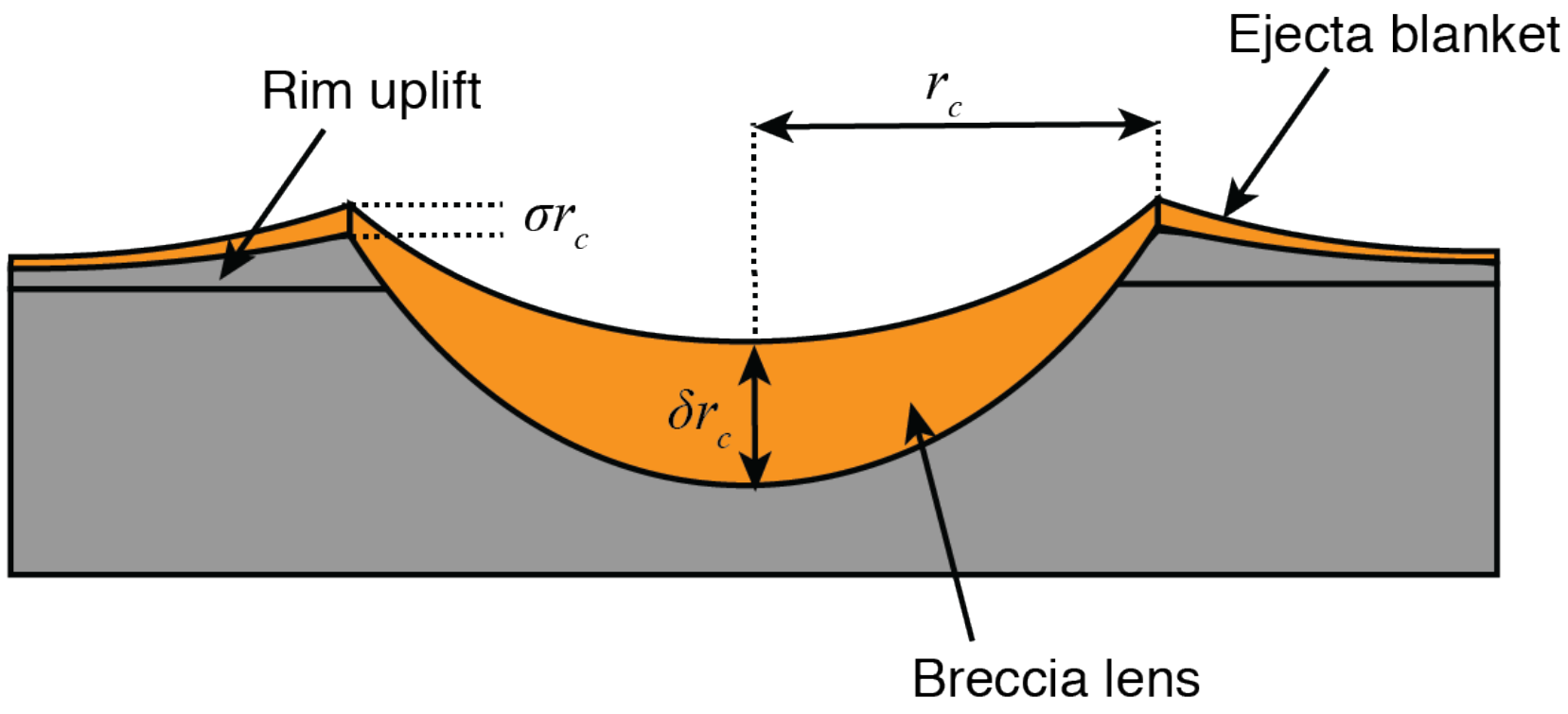

**Figure 1**. Schematic plot of the regolith region after a simple crater is formed. The gray region is a non-regolith region, while the orange area is a regolith region.

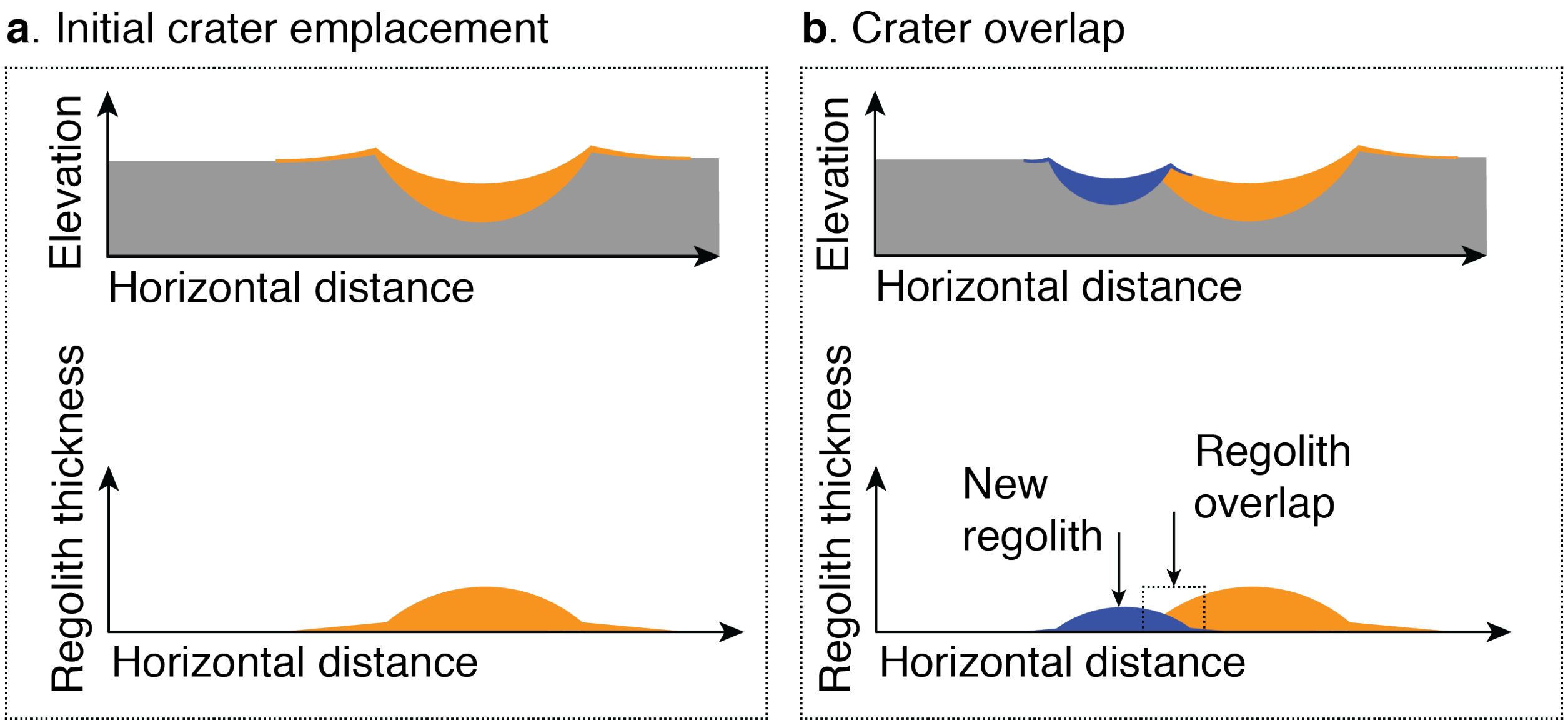

**Figure 2**. Schematic plot about how the analytical model deals with crater overlapping. The top figure shows the elevation along the horizontal direction. The gray area is the intact region. The bottom plot indicates the regolith thickness along the horizontal direction. a. Initial crater formation. b. Crater overlap. The orange region is the regolith area produced by the initial crater, while the dark blue region is the regolith region developed by a new crater's formation.

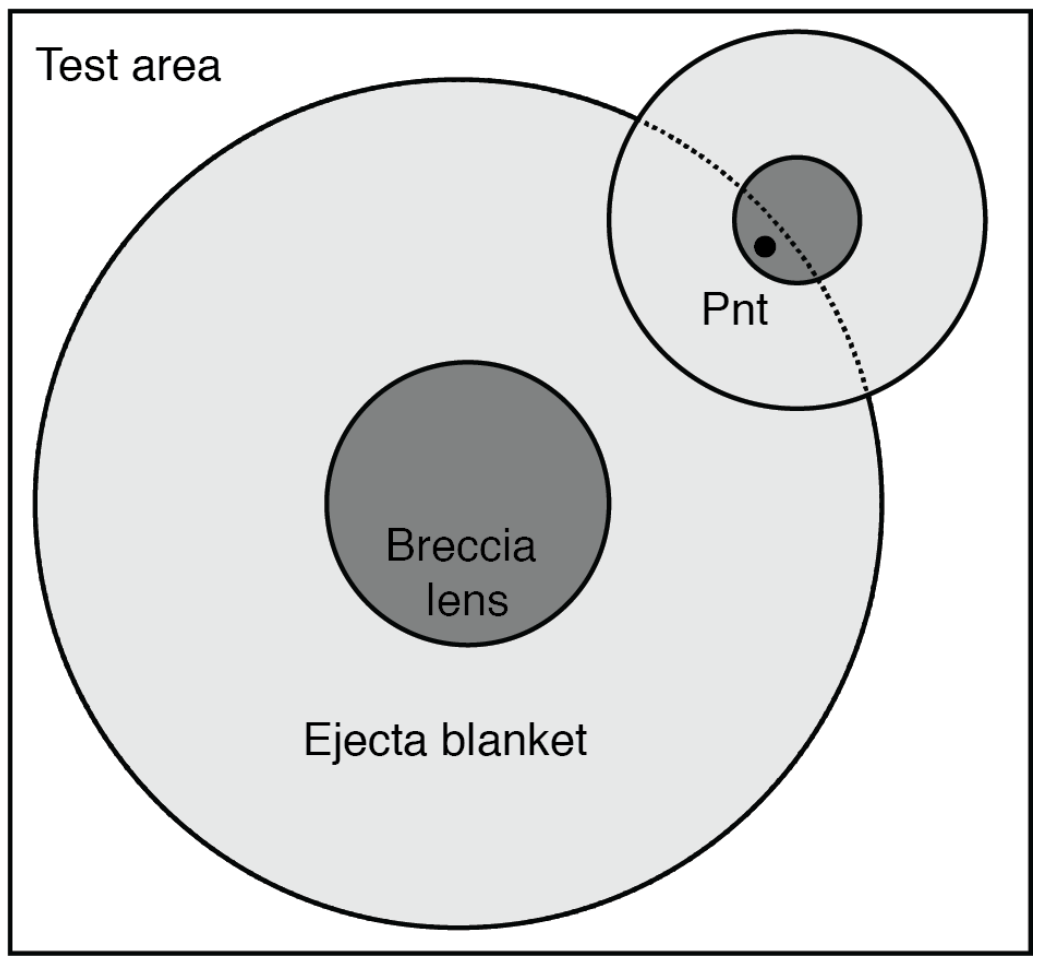

**Figure 3**. Schematic plot that explains the maximum crater radius contributing to regolith production due to regolith infilling on the transient crater cavity and that by ejecta blanketing. The dark gray regions are the breccia lens regions, while the light gray areas are the ejecta-blanket regions.

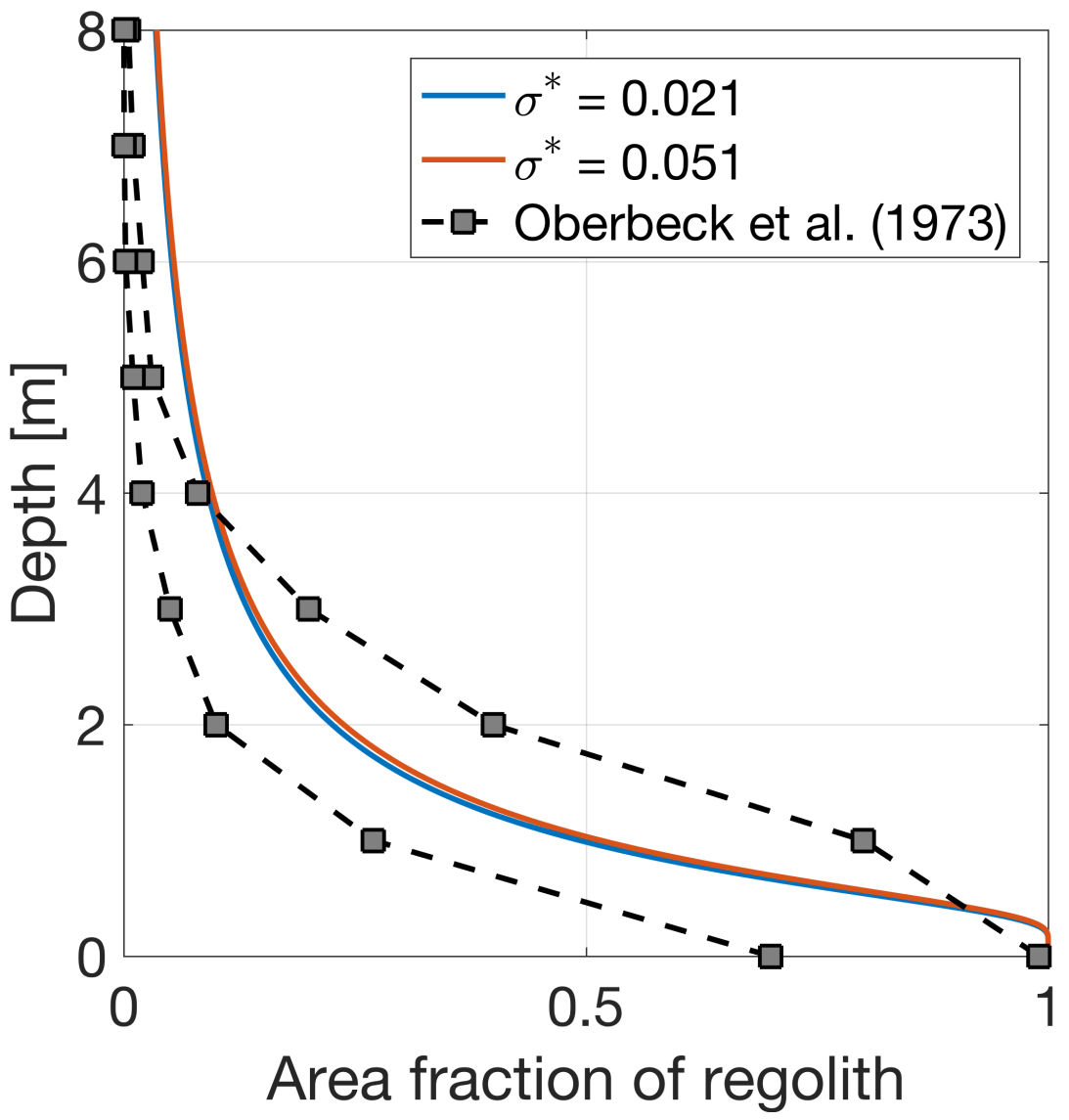

**Figure 4**. Comparison of the analytical model with the Monte Carlo simulation by *Oberbeck et al.* [1973]. The black dashed lines with squares describe the range of the results in Figure 1a in *Oberbeck et al.* [1973]. The blue and red lines show the results of $\sigma^* = 0.021$ and $\sigma^* = 0.051$, respectively, obtained by the analytical model.

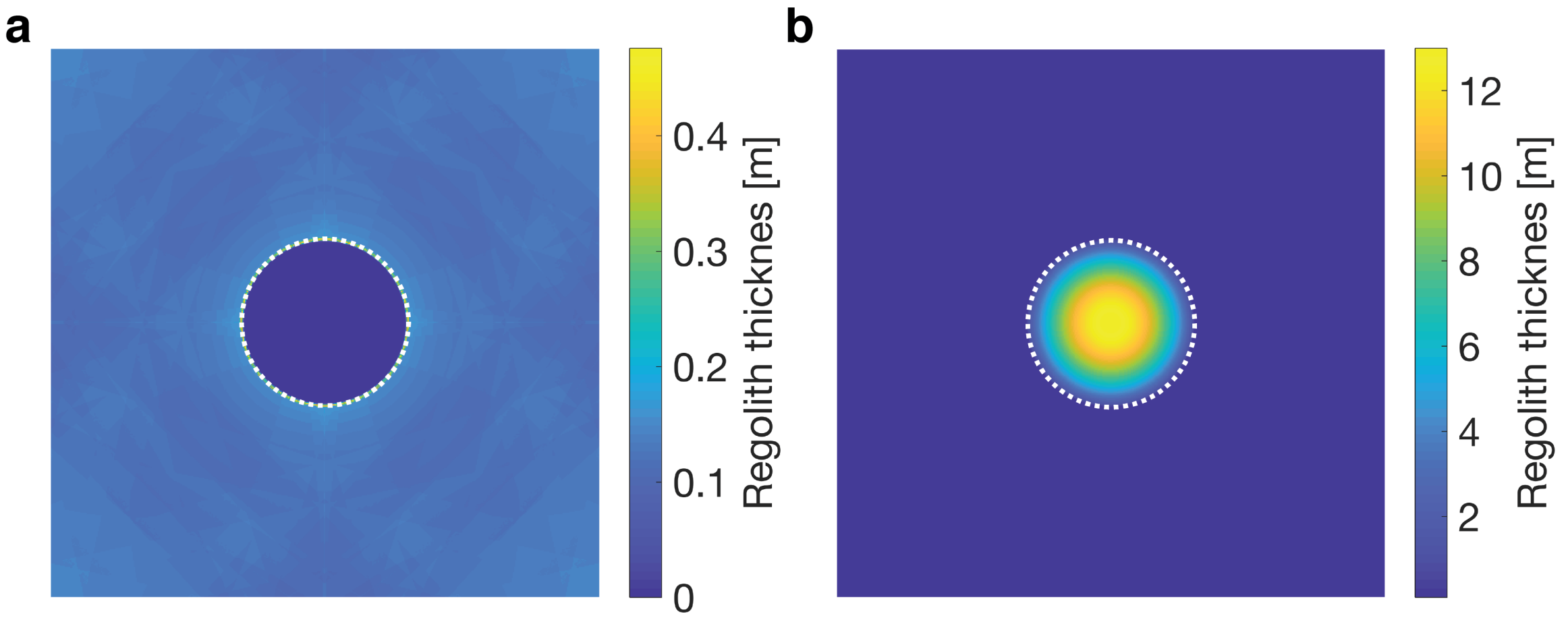

**Figure 5**. Regolith thickness by CTEM. The crater radius in this case is 38.2 m. a. The case only accounts for an ejecta blanket. b. The case considers an ejecta blanket and a breccia lens. The white dashed circles indicate the crater rim.

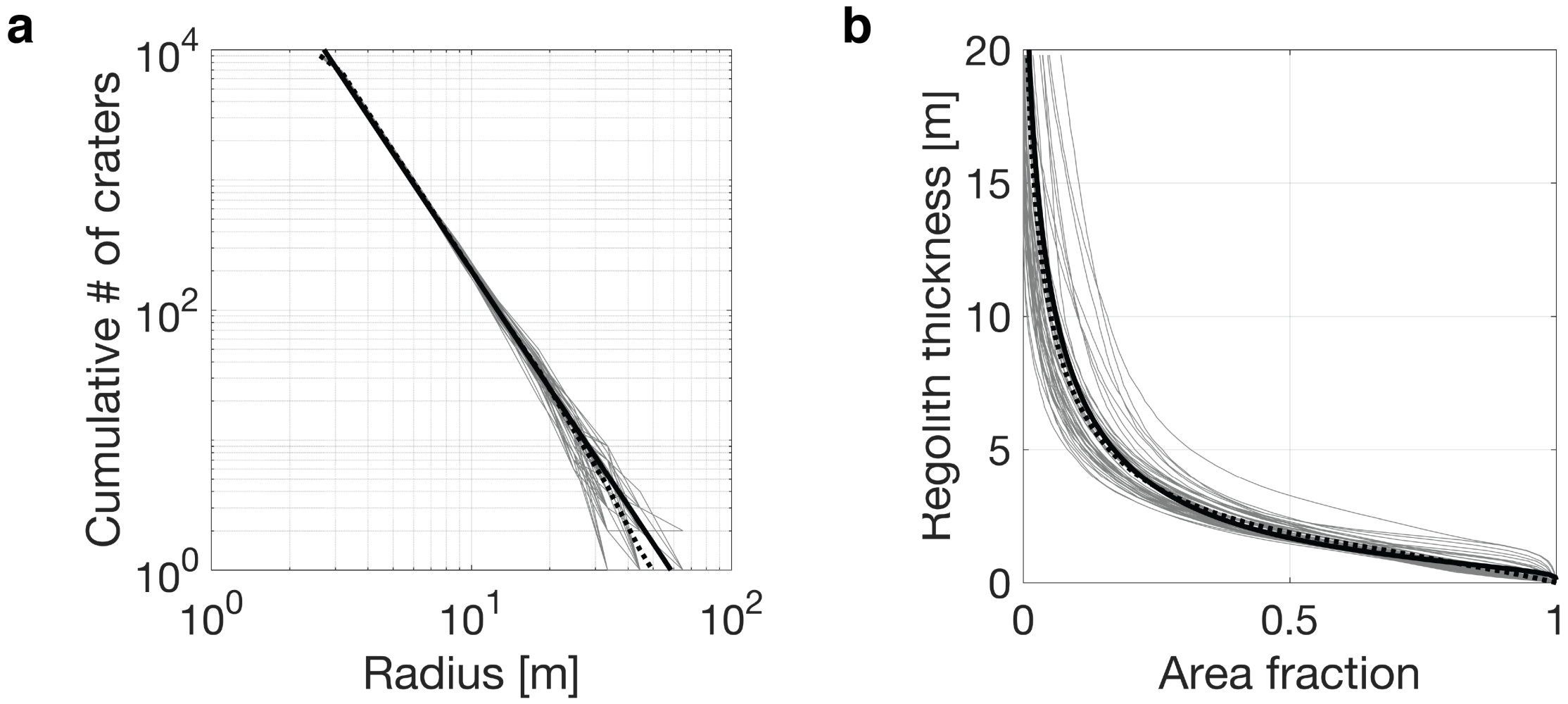

**Figure 6**. Comparison of the analytical model with CTEM. The gray solid lines show 50 cases computed by CTEM with different random seeds. The black solid lines indicate the result of the analytical model, and the black dashed lines describe the means of the results from CTEM. a. The produced crater CTEM. b. The regolith area fraction. The $x$ axis is the area fraction, while the $y$ axis shows the depth.

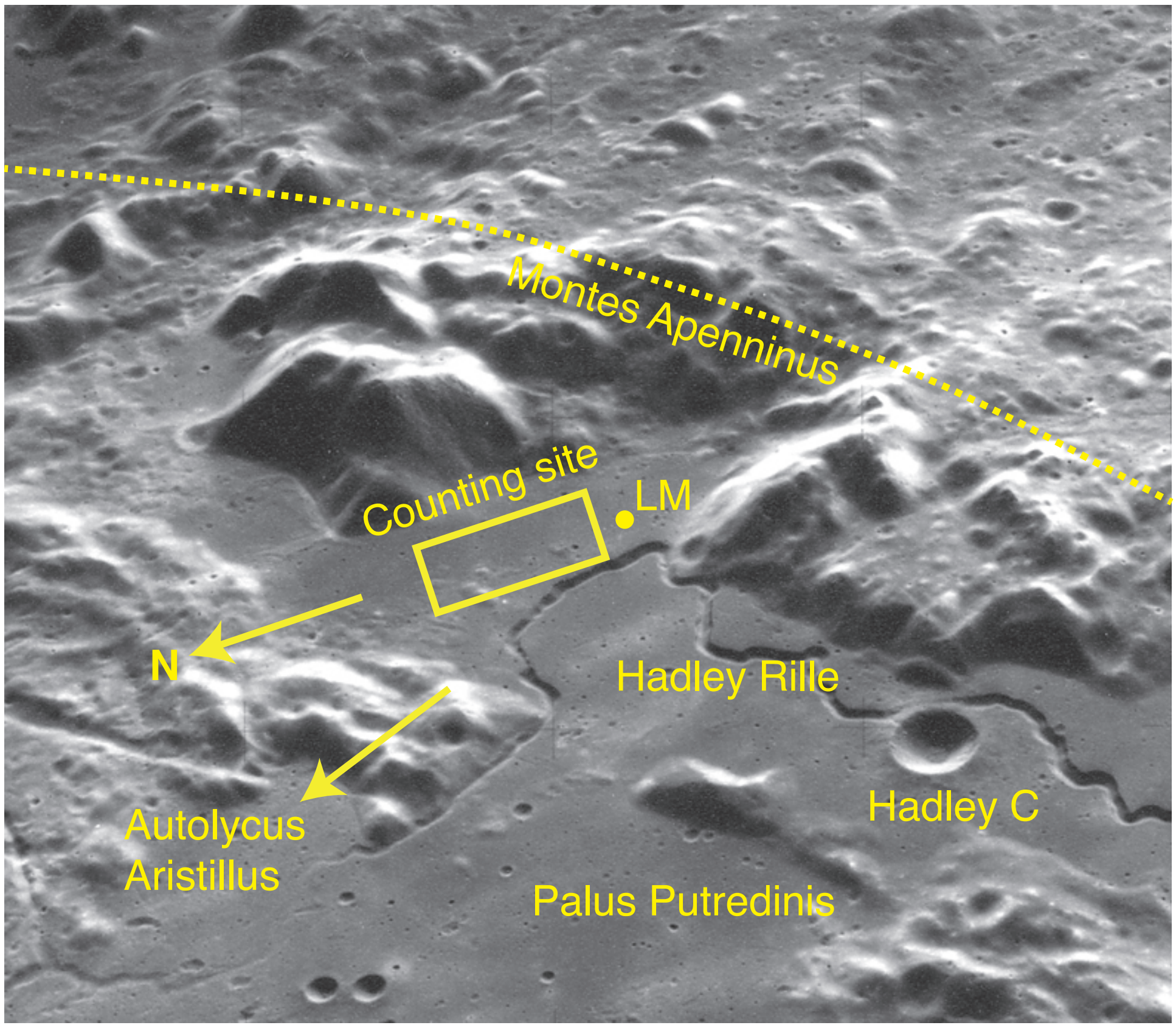

**Figure 7**. The landscape around the Apollo 15 site. Reference of the image: AS15-M-1423.

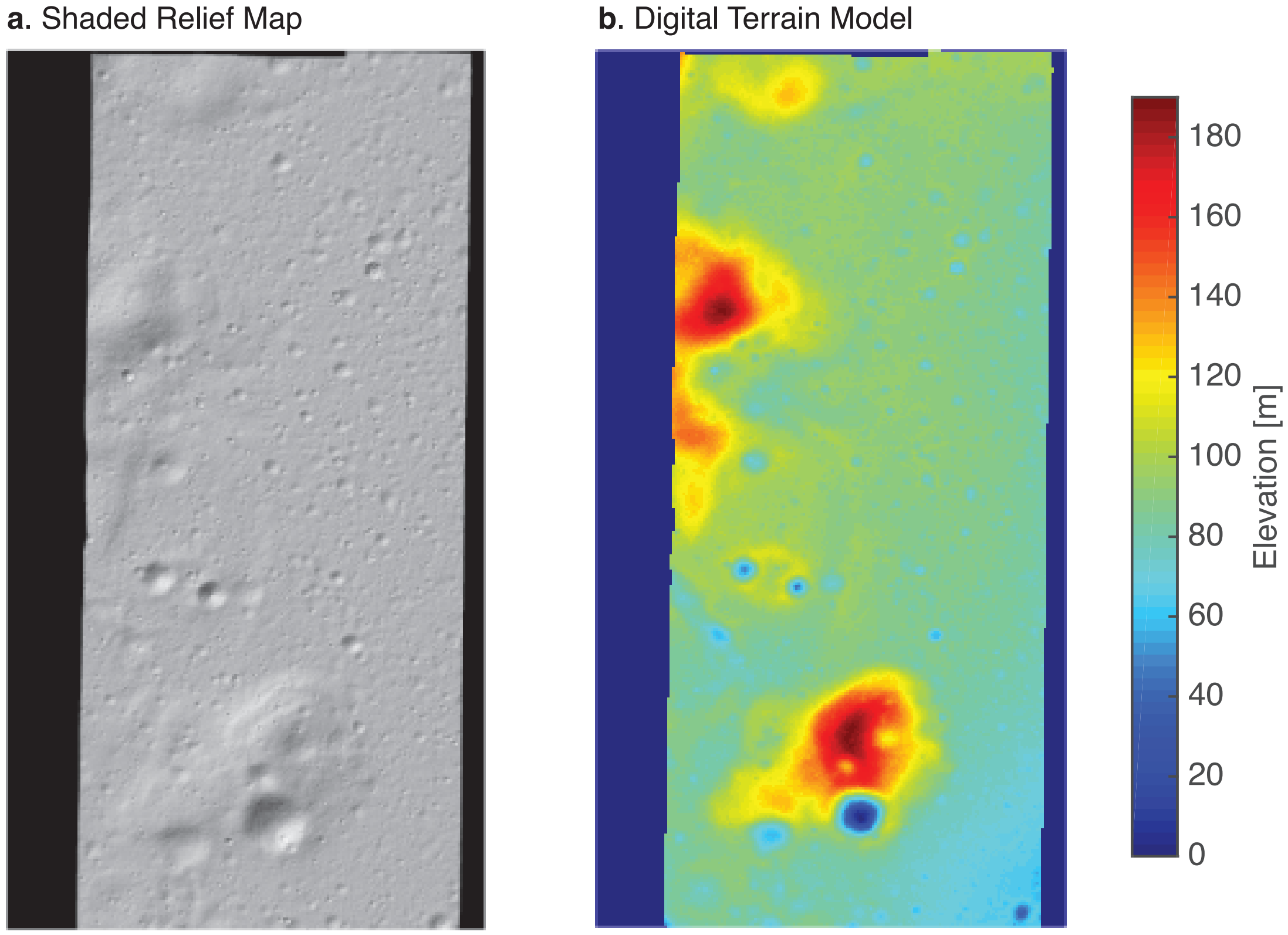

**Figure 8**. The region that we conduct crater counting. The selected area is the northern region in the original image (25.59–26.54°N, 3.50–3.69°E). a. The shaded relief map of the selected area (NAC_DTM_APOLLO15_SHADE.TIF). b. The DTM of the selected area (NAC_DTM_APOLLO15.TIF).

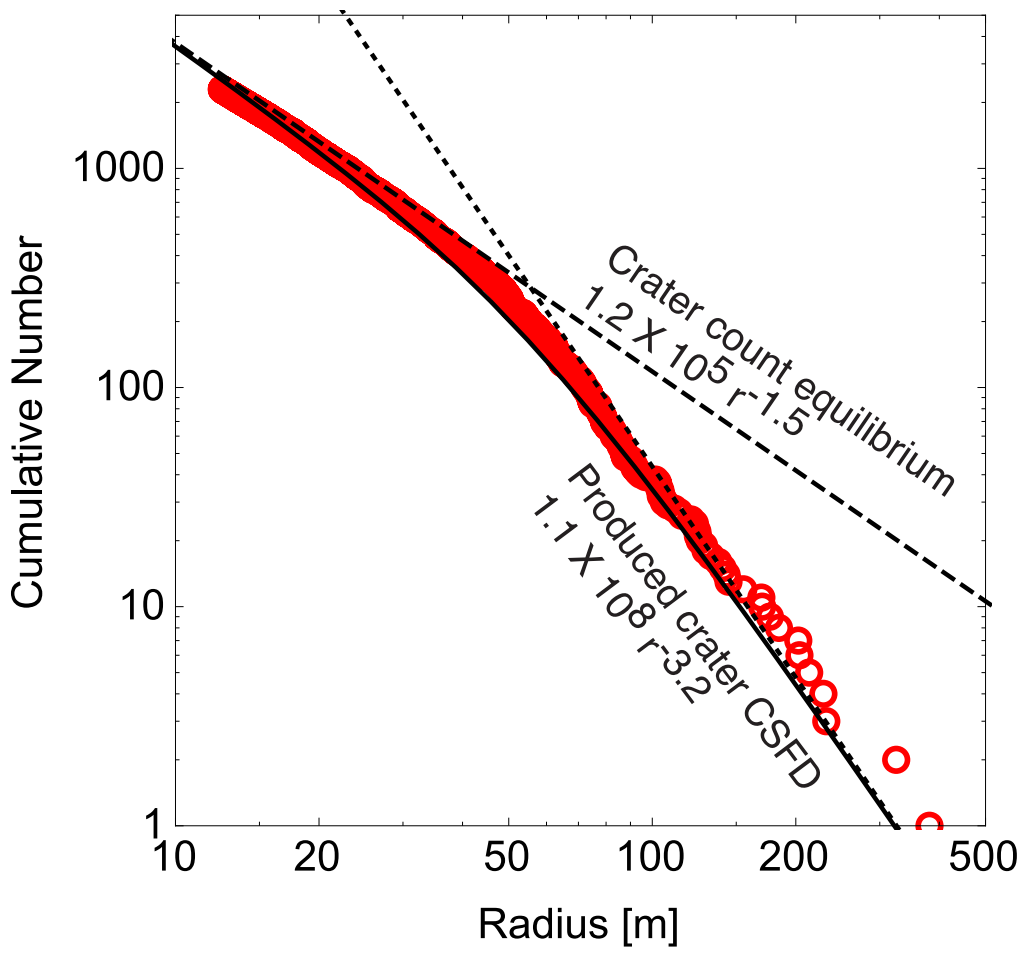

**Figure 9**. The visible crater CSFD at the Apollo 15 site in log-log space. The red-edge circles indicate our empirical result by crater counting. The solid line describes the best fitting function obtained by the technique by *Hirabayashi et al.* [2017]. The equilibrium slope is given by the dashed line, and the produced crater CSFD is described by the dotted line.

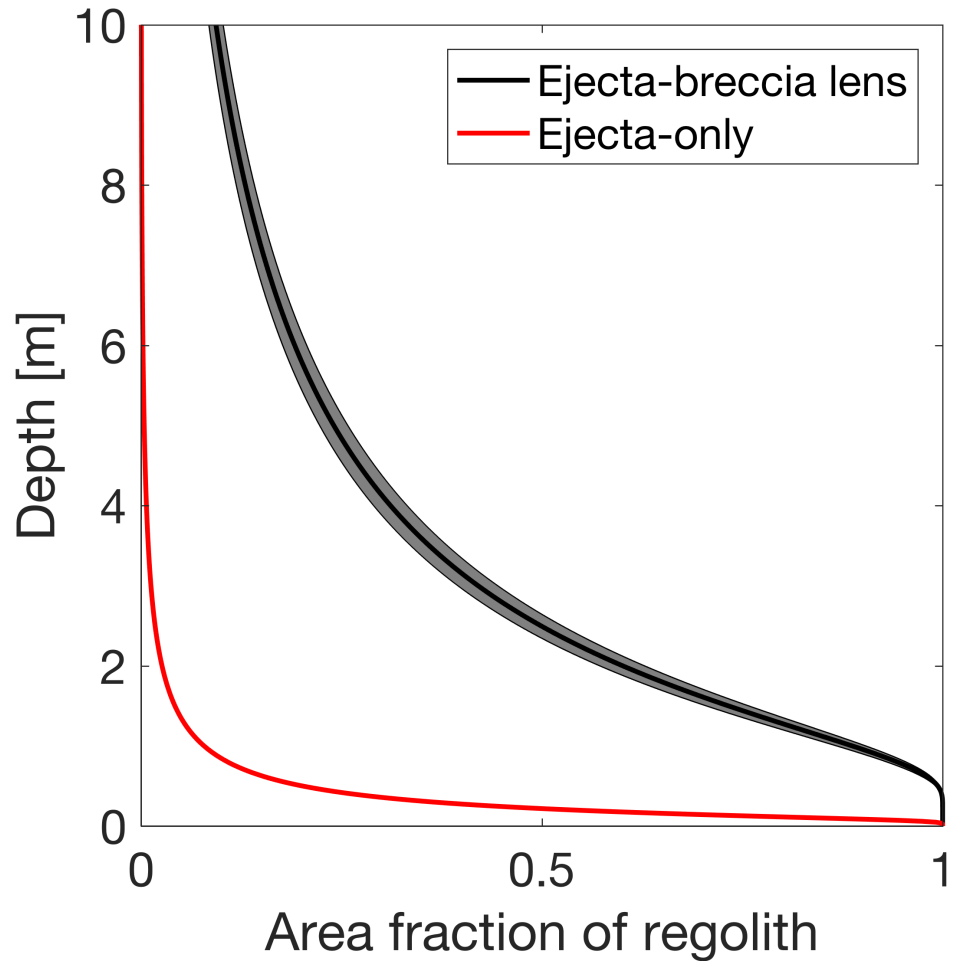

**Figure 10**. The regolith area fraction at the Apollo 15 landing site. The $x$ axis is the area fraction, while the $y$ axis shows the depth. The black line with a gray area describes the case considering an ejecta blanket and a breccia lens. The results are obtained using $\delta$ raging between 0.30 and 0.34. The solid line in the middle of the gray area is the result for $\delta = 0.32$. The red line indicates the case only accounting for an ejecta blanket. Because the ejecta-only case is independent of $\delta$, it is uniquely determined.

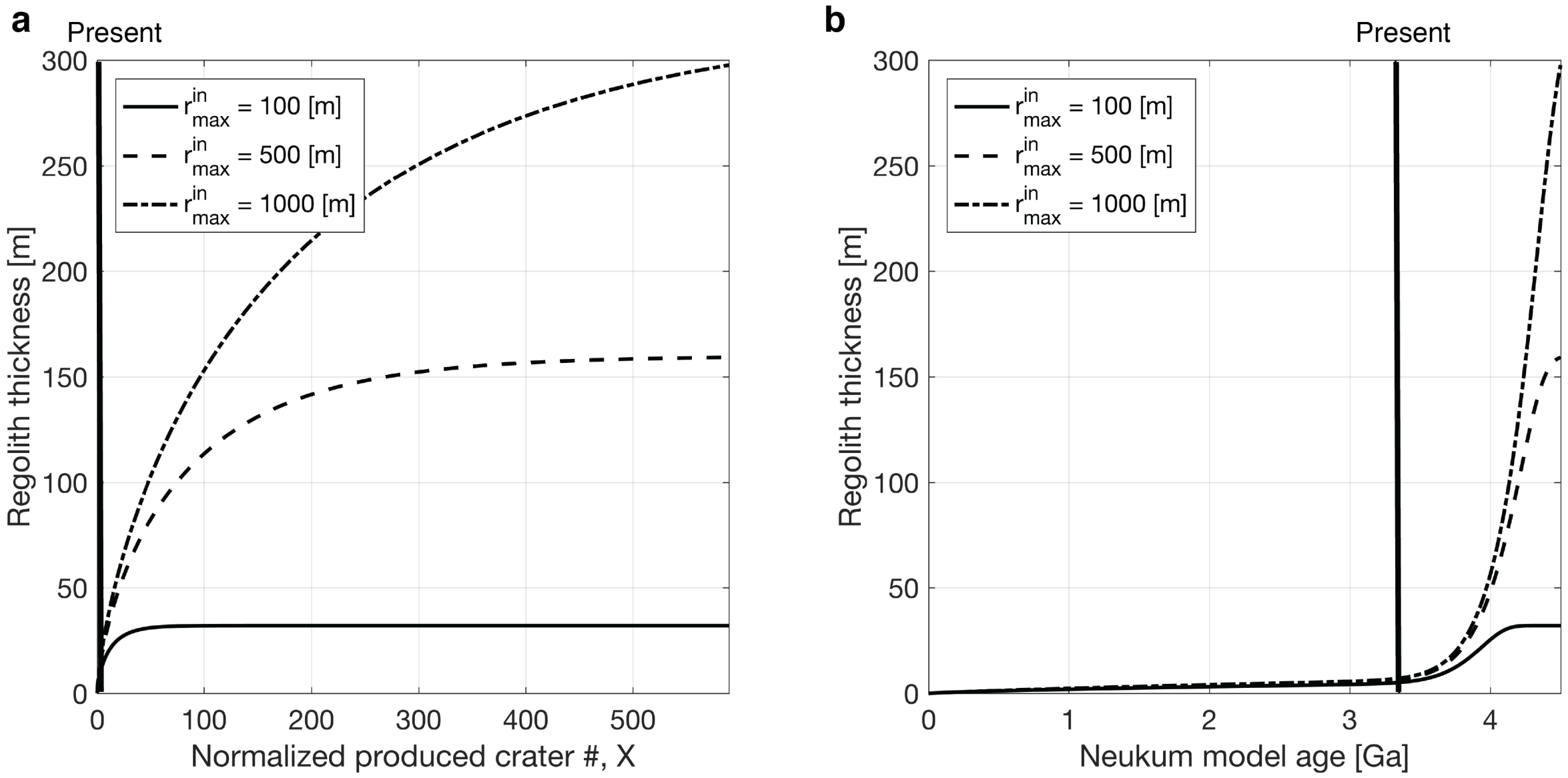

**Figure 11**. Evolution of the expected regolith thickness. a. The thickness evolution as a function of $X$ $(0 \leq X \leq 590)$. b. The thickness evolution as a function of the model age ($\leq$ 4.5 Ga).

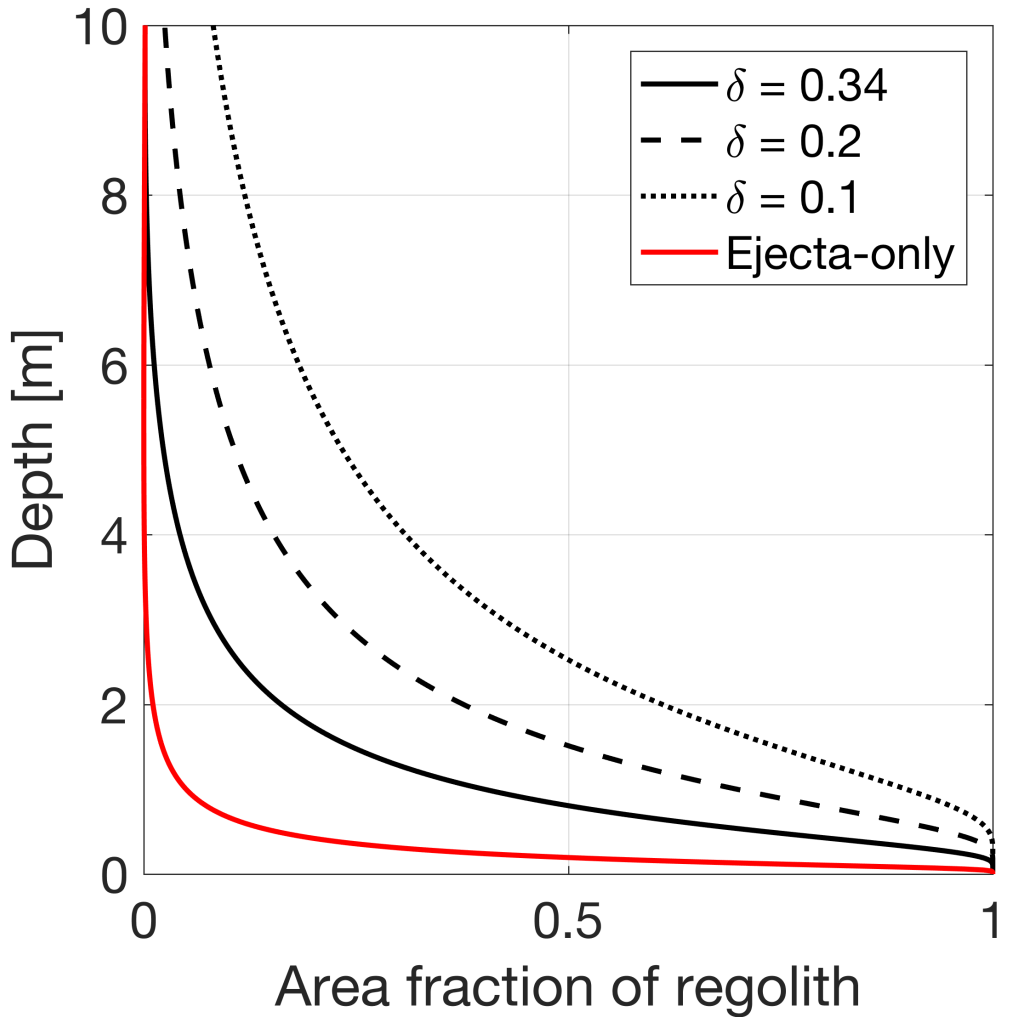

**Figure 12**. Regolith area fractions with different $\delta$s. The red line shows the ejecta-only case for comparison.

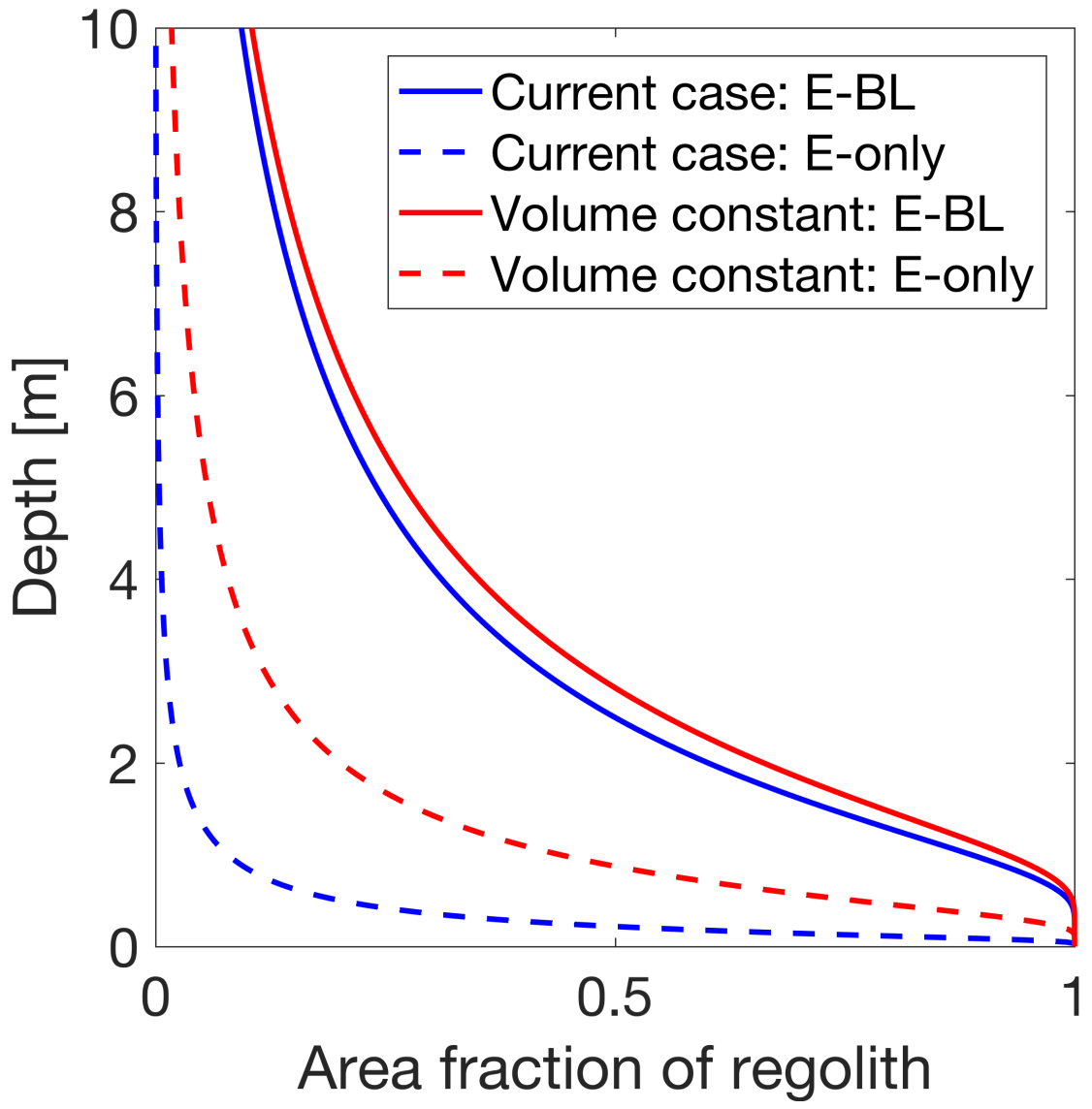

**Figure 13**. Comparison of the case derived using the parameters in Table 2 (in blue) and the case when the total crater volume is conserved (in red). "E-BL" is the ejecta-breccia lens case, while "E-only" shows the ejecta-only case. The solid lines are the ejecta-breccia lens cases, while the dashed lines are the ejecta-only cases.